\documentclass[reprint,nobibnotes,superscriptaddress]{revtex4-1}
\usepackage{epsf}
\usepackage{epsfig}
\usepackage{amsmath}
\usepackage{amsfonts}
\usepackage{amssymb}
\usepackage{graphicx}

\usepackage{lipsum}
\usepackage{nameref}
\usepackage{dsfont}

\usepackage{cleveref}
\usepackage[export]{adjustbox}
\usepackage{color}
\usepackage{natbib}

\usepackage{graphicx}
\usepackage{color}
\usepackage{bm}
\usepackage[yyyymmdd]{datetime}
\usepackage{dashrule}
\usepackage{hyperref}
\hypersetup{
    colorlinks=true,
    linkcolor=blue,
    filecolor=magenta,      
    urlcolor=red,
    citecolor=blue,
}
\usepackage{ulem}
\usepackage{todonotes}
\usepackage{xr}

\usepackage{cleveref}
\usepackage[export]{adjustbox}
\usepackage{color}

\newcommand{\be}{\begin{equation}}
\newcommand{\ee}{\end{equation}}

\def\bea{\begin{eqnarray}}
\def\eea{\end{eqnarray}}

\begin{document}

\title{Unraveling higher-order corrections in the spin dynamics of RIXS spectra}

\author{Umesh Kumar}
\email{umesh.kumar@rutgers.edu}
\affiliation{Theoretical Division, Los Alamos National Laboratory, Los Alamos, New Mexico 87545, USA}

\author{Abhishek Nag}
\email{abhishek.nag@diamond.ac.uk}
\affiliation{Diamond Light Source, Harwell Campus, Didcot OX11 0DE, United Kingdom}

\author{Jiemin Li}
\affiliation{Diamond Light Source, Harwell Campus, Didcot OX11 0DE, United Kingdom}
\affiliation{Beijing National Laboratory for Condensed Matter Physics and Institute of Physics, Chinese Academy of Sciences, Beijing 100190, China}

\author{H. C. Robarts} 
\affiliation{Diamond Light Source, Harwell Campus, Didcot OX11 0DE, United Kingdom}
\affiliation{H. H. Wills Physics Laboratory, University of Bristol, Bristol BS8 1TL, United Kingdom}

\author{A. C. Walters}
\affiliation{Diamond Light Source, Harwell Campus, Didcot OX11 0DE, United Kingdom}

\author{Mirian Garc\'ia-Fern\'andez}
\affiliation{Diamond Light Source, Harwell Campus, Didcot OX11 0DE, United Kingdom}

\author{R. Saint-Martin}
\affiliation{Institut de Chimie Mol\'eculaire et des Mat\'eriaux
d’Orsay, Universit\'e Paris-Saclay, UMR 8182, 91405 Orsay, France}

\author{A. Revcolevschi}
\affiliation{Institut de Chimie Mol\'eculaire et des Mat\'eriaux
d’Orsay, Universit\'e Paris-Saclay, UMR 8182, 91405 Orsay, France}

\author{Justine Schlappa}
\affiliation{European X-Ray Free-Electron Laser Facility GmbH, Holzkoppel 4, 22869 Schenefeld, Germany}

\author{Thorsten Schmitt}
\affiliation{Photon Science Division, Paul Scherrer Institut, 5232 Villigen PSI, Switzerland}

\author{Steve Johnston}
\affiliation{Department of Physics and Astronomy, The University of Tennessee, Knoxville, TN 37996, USA}

\author{Ke-Jin Zhou}
\email{kejin.zhou@diamond.ac.uk}
\affiliation{Diamond Light Source, Harwell Campus, Didcot OX11 0DE, United Kingdom}

\date{\today}

\begin{abstract}
Resonant inelastic x-ray scattering (RIXS) is an evolving tool for investigating spin dynamics of strongly correlated materials, which complements inelastic neutron scattering. Both techniques have found that non-spin-conserving (NSC) excitations in quasi-1D isotropic quantum antiferromagnets are confined to the two-spinon phase space. Outside this phase space, only spin-conserving (SC) four-spinon excitations have been detected using O $K$-edge RIXS. Here, we investigate SrCuO$_2$ and find four-spinon excitations outside the two-spinon phase space at both O $K$- and Cu $L_3$-edges. Using the Kramers-Heisenberg formalism, we demonstrate that the four-spinon excitations arise from both SC and NSC processes at Cu $L_3$-edge. We show that these new excitations only appear in the second-order terms of the ultra-fast core-hole lifetime expansion and arise from long-range spin fluctuations. These results thus open a new window to the spin dynamics of quantum magnets.
\end{abstract}

\maketitle
The elementary excitations of a system encode information about the microscopic origin of its behavior. For example, magnetic excitations are extensively studied in high-$T_c$  superconducting cuprates for their possible role in the pairing mechanism and provide access to critical physical parameters like their correlation strength, superexchange interactions, or electronic structure parameters~\cite{SCALAPINO1995329, Chubukov2003, Orenstein468, Sidis2004, Scalapino2012, Lee2006, Headings2010}. Inelastic neutron scattering (INS) has traditionally been the method of choice for probing magnetic correlations. In recent years, however, resonant inelastic x-ray scattering (RIXS) has emerged as an important complementary technique to INS~\cite{RevModPhys.83.705,PhysRevX.6.021020,PhysRevLett.104.077002,PhysRevLett.114.217003,LeTacon2011,Dean2012,PhysRevLett.108.177003,PhysRevX.8.031014,HCRobarts_2021,PhysRevLett.124.067202,PhysRevLett.103.117003, PhysRevLett.104.077002, RevModPhys.93.035001}. Its large scattering cross-section and element specificity are advantageous over INS for small sample volumes, as well as for complex systems with multiple magnetic elements~\cite{pelliciari2021evolution, PhysRevB.99.134414}. 

While the relationship between the INS cross-section and the  dynamical spin structure factor $S({\bf q},\omega)$ is well understood, the same cannot be said about the RIXS cross-section, which is described by the Kramers-Heisenberg (KH) formalism~\cite{RevModPhys.83.705}. Here, the presence of a core-hole in the intermediate state allows for a diverse set of scattering channels including both spin-conserving (SC, $\Delta S_{\mathrm{tot}}=0$) and non-spin-conserving (NSC, $\Delta S_{\mathrm{tot}}=1$) processes. The RIXS intensity can only be directly equated with the dynamical structure factors under certain scattering conditions \cite{Jia2014} or in the limit of ultra-short core-hole lifetimes \cite{PhysRevB.75.115118}, a fact that is becoming more evident with the recent developments in resolving the polarization of the scattered x-rays~\cite{betto2021multiplemagnon, PhysRevLett.114.217003, PhysRevB.99.134517}. While this aspect complicates the interpretation of RIXS experiments, it has also led to the observation of novel excitations such as orbitons~\cite{Nature.485.7396, PhysRevB.88.195138} and SC four-spinon excitations outside two-spinon phase space~\cite{FSDIR, KUMARNJP2018} in the one-dimensional (1D) Heisenberg antiferromagnetic (HAFM) spin chain Sr$_2$CuO$_3$.


The relationship between the RIXS intensity and more traditional dynamical structure factors can be derived systematically using the ultra-short core-hole lifetime (UCL) expansion~\cite{Jia2014, PhysRevB.75.115118, PhysRevB.83.245133, PhysRevX.6.021020}. Here, one expands the RIXS intensity as a series of increasingly complicated multi-particle correlation functions in powers of the inverse core-hole lifetime $1/\Gamma$. For example, the transition metal $L$-edge RIXS spectra of a $3d^9$ system can be approximated as $I({\bf q},\omega) \approx |W|^2\left[S({\bf q}, \omega)/\Gamma + S^\prime({\bf q}, \omega)/\Gamma^2\right]$ in certain scattering conditions \cite{PhysRevX.6.021020}. Here, $W$ is a matrix element related to the scattering geometry, $S({\bf q}, \omega)$ is the dynamical spin structure factor, and $S^\prime({\bf q}, \omega)$ is the leading correction. Notably, the quantities appearing in the expansion can be computed in the absence of the core-hole, allowing one to use numerous methods to attack the problem. For example, the spectra for the SC channel generated via double spin-flips in 1D cuprate CaCu$_2$O$_3$~\cite{PhysRevLett.112.147401} were shown to fractionalize entirely into two-spinons using the Bethe ansatz~\cite{PhysRevLett.106.157205}. In the case of spin-ladders, three-triplon bound states were predicted in the same channel~\cite{Ferkinghoff2020} using the continuous unitary transformation method. However, the extent to which higher-order corrections to the UCL expansion are required for the NSC excitations at $L$-edges~\cite{PhysRevX.6.021020, UKUMAR2019, PhysRevB.77.134428} has not been widely studied.

In this letter, we report the spin dynamics of the 1D HAFM SrCuO$_2$ probed by high-resolution Cu $L_3$- and O $K$-edge RIXS. We observe a two-spinon continuum, in agreement with INS on this system~\cite{PhysRevLett.93.087202}. Unlike INS, however, we also observe considerable spectral weight from four-spinon excitations located in a region of phase space outside of the two-spinon continuum.  Moreover, the distribution of spectral weight for these excitations observed at each edge is distinct. The four-spinon excitations observed at the O $K$-edge arise from the SC channel as found in the related work on Sr$_2$CuO$_3$~\cite{FSDIR}. Using detailed theoretical modeling within the KH formalism for the $t$-$J$ model, we demonstrate that the four-spinon excitations observed at Cu $L_3$-edge spectra arise from both SC and NSC scattering channels. We further evaluate new dynamical correlation functions derived from the UCL expansion to show that these four-spinon excitations can only be accounted for by including second-order terms of the expansion involving long-range spin excitations. These observations thus open up possibilities for investigating long-range many-body excitations in quantum magnets.

\begin{figure}[tbp]
	\begin{center}
		\includegraphics[width=84mm]{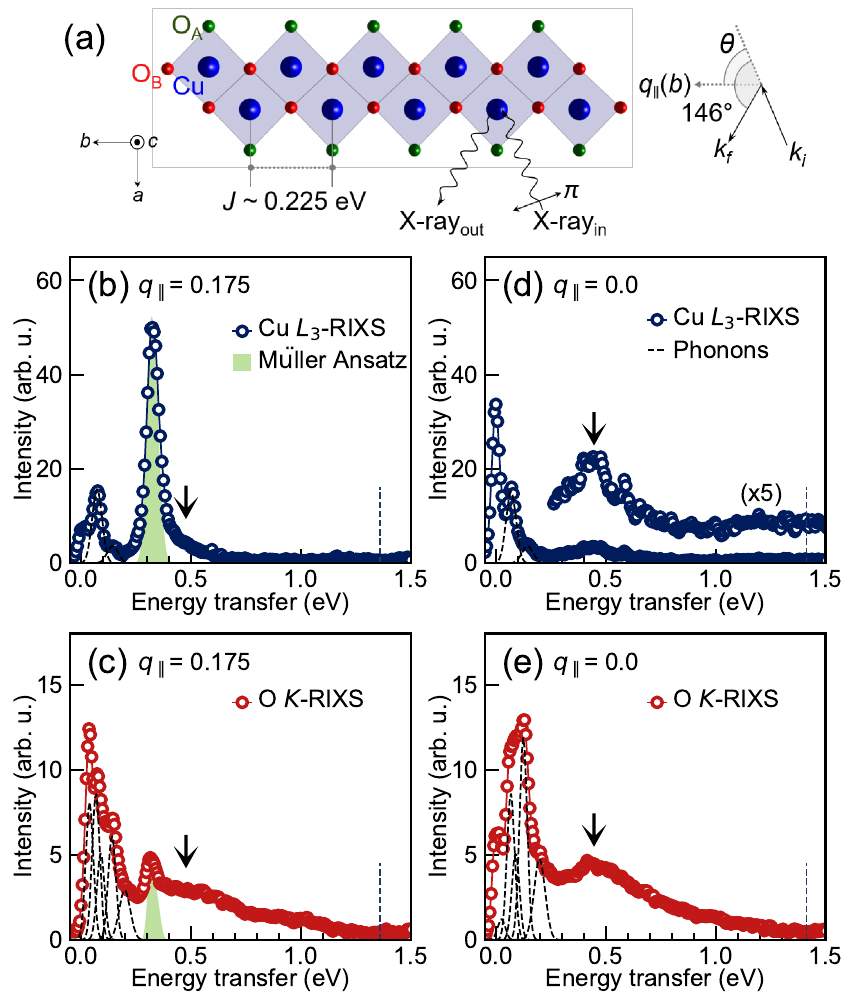}	%
		\caption{(a) A schematic diagram of the RIXS scattering geometry for SrCuO$_2$. Panels (b) and (c) show the Cu low-energy Cu $L_3$- and O $K$-edge RIXS spectra, respectively, measured at $q_\parallel= 0.175$. Panels (d) and (e) show the corresponding data at $q_\parallel = 0$. Also shown in (d) is the RIXS spectrum multiplied by a factor of five and vertically shifted. At $q_{\parallel}=0$, the energy of the two-spinon excitations and their spectral weight are zero. For $q_{\parallel}=0.175$, the two-spinons are approximated by the M$\ddot{\mathrm{u}}$ller Ansatz~\cite{PhysRevB.55.12510}. The vertical dashed line indicates the upper boundary of the four-spinon continuum~\cite{1742-5468-2006-12-P12013}. In all cases, additional spectral weight (marked by arrows) is observed outside the two-spinon phase space but inside the four-spinon phase space.
		} \label{fig1} 
	\end{center}
	\vspace{-0.75cm}
\end{figure}

SrCuO$_2$ contains edge-shared CuO$_4$ plaquettes arranged in a zig-zag geometry  with negligible inter-chain coupling [Fig.~\ref{fig1}(a)]~\cite{PhysRevLett.76.3212,PhysRevLett.77.4054, Kim2006,PhysRevLett.93.087202,PhysRevB.55.R7291,PhysRevLett.83.5370}. The Cu sites have $3d^9$ valence configurations in the atomic limit with a hole in the $d_{x^2-y^2}$ orbital. Due to strong intra-chain coupling, the system effectively behaves as a 1D spin-$\frac{1}{2}$ HAFM chain above $T_N=5$~K~\cite{PhysRevLett.93.087202, PhysRevLett.83.5370}. The zig-zag arrangement of the plaquettes creates two inequivalent oxygen sites denoted O$_A$ and O$_B$, as shown in Fig.~\ref{fig1}(a). Due to the element specificity of the RIXS process, excitations of the in-chain O$_B$ or Cu sites can be isolated by fixing the incident energy to X-ray absorption spectra (XAS) peaks from these respective elements. We obtained the resonant energies of the O$_B$ and Cu sites from the XAS collected at O $K$- and Cu $L_3$-edges, respectively (see Supplementary Materials Fig. S1)~\cite{PhysRevB.55.R7291,SM}. RIXS spectra on SrCuO$_2$ at 13~K were collected at O $K$- and Cu $L_3$-edges with energy resolutions of $\sim 27$ and $\sim 42$ meV, respectively, at I21 beamline, Diamond Light Source, United Kingdom~\cite{I21}. Throughout, the momentum transfer component along the chain direction ($q_{\parallel}$) is presented in units of 2$\pi/b$.

Figures~\ref{fig1}(b) and (c) show the RIXS spectra from SrCuO$_2$ for $q_{\parallel}=0.175$ at the Cu $L_3$- and O $K$-edges, respectively. Similar to RIXS results on the related compound Sr$_2$CuO$_3$, phonon excitations and their overtones are observed below $\sim$0.2~eV for both edges~\cite{FSDIR, Nature.485.7396}. The intense feature close to 0.35~eV energy loss at the Cu $L_3$-edge can be ascribed to a two-spinon excitation and approximated by the dynamical structure factor computed using the M{\"u}ller ansatz after, adopting an exchange coupling $J = 0.225$~eV~\cite{PhysRevB.55.12510, PhysRevLett.93.087202}. A weaker feature is also observed at O $K$-edge at a similar energy. Panels (d) and (e) show the RIXS spectra for $q_{\parallel}=0.0$ at Cu $L_3$- and O $K$-edges, respectively. For both edges, significant spectral weight is observed, peaking close to 0.5~eV. Since the weight of the two-spinon continuum is zero at this momentum point, and the orbital and charge-transfer excitations lie above 1.5~eV, these features likely arise from four-spinon excitations~\cite{FSDIR}. Overall, it appears that RIXS probes similar excitations  at both edges but with substantially different scattering cross-sections, giving rise to distinct spectral weight distributions. For a better understanding of the RIXS cross-sections, we next map these excitations in the energy-momentum space.

\begin{figure}[tb]
	\begin{center}
		\includegraphics[width=89mm]{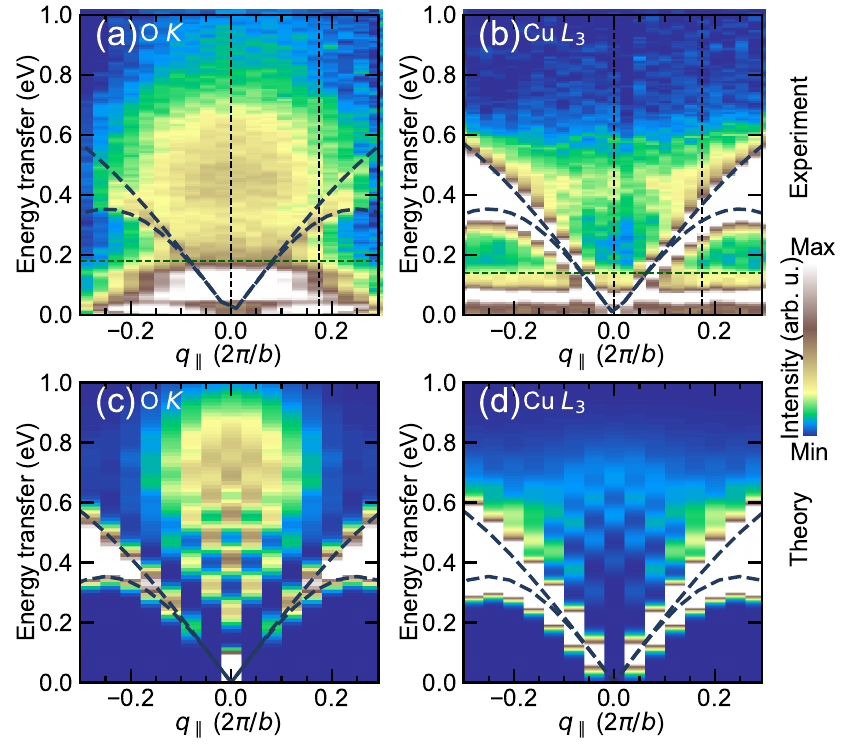}	%
		\caption{(a) O $K$-edge and (b) Cu $L_3$-edge RIXS intensity maps for momentum transfer along the chain. The white colour represents the maximum spectral weight in the colour scale. The two-spinon continuum boundaries are shown by blue dashed lines. Horizontal dashed green lines mark the higher energy limit of the phonon excitations. The RIXS spectra shown in Fig.~\ref{fig1}(b-e) correspond to the the vertical black dashed lines. (c) Model intensity maps for spin excitations at the (c) O $K$- and (d) Cu $L_3$-edge computed using the Kramers-Heisenberg formalism.} 
		\label{fig2}
	\end{center}
	\vspace{-0.75cm}
\end{figure}

Figures~\ref{fig2}(a) and (b) present the momentum resolved RIXS intensity maps at the O $K$- and Cu $L_3$-edges, respectively. The lower and upper boundaries of the two-spinon phase space, $\omega_\mathrm{2S}^\mathrm{l} (\textbf{q}) = \frac{\pi}{2} J |\sin(qb)|$ and $\omega_\mathrm{2S}^\mathrm{u} (\textbf{q}) = \pi J |\sin(qb/2)| $~\cite{PhysRevB.55.12510}, are plotted over the maps for $J=0.225$~eV. A highly dispersing spectral weight is observed within this phase space, consistent with INS experiments on SrCuO$_2$~\cite{PhysRevLett.93.087202}. The two-spinon continuum composed of fractionalized spin-1/2 entities has been previously observed at Cu $L_3$-edge RIXS for Sr$_2$CuO$_3$, Ca$_2$CuO$_3$ and CaCu$_2$O$_3$ ~\cite{Nature.485.7396, FSDIR, PhysRevB.101.205117, PhysRevLett.112.147401}, where they arise from a  spin-flip scattering process enabled by the strong spin-orbit coupling in the $2p$ core level. In O $K$-edge RIXS, where a single spin-flip is not allowed, two-spinon excitations are created by a net spin-zero inter-site double spin-flip process~\cite{PhysRevB.83.245133, FSDIR}. Further, due to the longer core-hole lifetime at O $K$-edge, quantum fluctuations lead to formation of additional domain walls in the intermediate state~\cite{FSDIR}. Four-spinon excitations are thus formed, appear as an additional continuum of excitations located outside the two-spinon phase space [Fig~\ref{fig2}(a)] but within the four-spinon phase space extending up to $\omega_\mathrm{4S}^\mathrm{u} (\textbf{q}) =\pi J\sqrt{2[1+\cos (qb/2)]} $ ($\sim1.4$ eV for $J=0.225$ close to $q_{\parallel}=0$)~\cite{1742-5468-2006-12-P12013}.

Interestingly, we also observe significant spectral weight outside the two-spinon phase space in Cu $L_3$-edge measurements. Similar features were reported by Ref.~\cite{FSDIR}, but were not explored further. Notably, the four-spinon spectral weight distributions are drastically different between the two edges for the same system. While at the O $K$-edge they forms a bulbous shape with largest weights close to $q_{\parallel}=0$, at Cu $L_3$-edge they have a flatter distribution with weight collected near the upper boundary of the two-spinon continuum. 

The low-energy spin dynamics of corner-shared cuprates are well described by $t-J$ Hamiltonian~\cite{Nocera2018, PhysRevLett.93.087202, Walters2009,FSDIR} 
\begin{equation}\label{eq:tJHamiltonian}
	H=-t\sum_{\langle i,j\rangle,\sigma}\tilde{c}^\dagger_{i,\sigma}\tilde{c}^{\phantom\dagger}_{j,\sigma}+ J\sum\limits_{i}\big(\textbf{S}_i\cdot \textbf{S}_{i+1} - \frac{1}{4} n_i n_{i+1}\big).  
\end{equation}
Here, $t$ and $J$ are hopping integrals and exchange coupling, respectively, between the nearest neighbours $\langle i, j\rangle$. $\tilde{c}^\dagger_{i,\sigma}~(\tilde{c}^{\phantom\dagger}_{i,\sigma})$ is the creation (annihilation) operator for a spin-$\sigma$ ($= \uparrow,\downarrow$) hole at site $i$ under the constraint of no double occupancy; $n_{i}=\sum_{\sigma}\tilde{c}^\dagger_{i,\sigma}\tilde{c}^{\phantom\dagger}_{j,\sigma}$ is the number operator; and $\textbf{S}_{i}$ is the spin operator at site $i$. We use $J=0.225$~eV from the energy-momentum distribution of the two-spinon continuum in our RIXS experiments and $t=0.3$, typical for cuprates. 

We solved Eq.~(\ref{eq:tJHamiltonian}) using exact diagonalization (ED) on twenty-four site chains and evaluated the RIXS intensity $I(q,\omega)$ using the KH formalism~\cite{RevModPhys.83.705}  
\begin{equation}
I(q,\omega) \propto \sum\limits_{f}\left\lvert
\sum\limits_{n}\frac{\langle f \lvert D^{\dagger}_{k_\mathrm{out}}\rvert n \rangle \langle n \lvert D^{\phantom\dagger}_{k_\mathrm{in}} \rvert g \rangle}
{E_{g} + \omega_\mathrm{in} - E_{n} + i \Gamma_{n}} \right\rvert^{2}
\delta\left(E_{f} - E_{g} - \omega\right).
\end{equation}
\noindent
In the above expression, the incoming (outgoing) photons  have energy $\omega_\mathrm{in}$ ($\omega_\mathrm{out}$) and momentum $k_\mathrm{in}$ ($k_\mathrm{out}$); 
$\omega = \omega_\mathrm{in} - \omega_\mathrm{out}$ and 
$q = k_\mathrm{in}-k_\mathrm{out}$ are the energy and momentum transferred along the chain direction, respectively; $\lvert g
\rangle$, $\lvert n \rangle$, and $\lvert f \rangle$ are the initial, intermediate, and final states of the RIXS process with energies $E_{g}$, $E_{n}$, and $E_{f}$, respectively; $D_{k}= \sum_{i, \sigma} e^{i k R_i} D_{i,\sigma}$ is the dipole operator describing the $2p\rightarrow 3d$ $\left(D_{i,\sigma} = \sum_{\alpha} d^{\phantom\dagger}_{i \sigma} p_{i\sigma}^\dagger\right)$ and $1s\rightarrow 2p$  $\left(D_{i,\sigma} = [d^{\phantom\dagger}_{i+1, \sigma}-d^{\phantom\dagger}_{i-1, \sigma}]  s_{i\sigma}^\dagger\right)$  transition~\cite{KUMARNJP2018, FSDIR}, where $p^\dagger_{i,\sigma}$ and $s_{i\alpha}^\dagger$ are creation operators on the $2p$ and $1s$ core levels, respectively; and $\Gamma_n$ is the core-hole broadening related to the inverse core-hole lifetime. For our numerical calculations we use $\Gamma_n = 0.3$ eV (Cu $L_3$-edge) and  $0.15$ eV (O $K$-edge) for all $n$~\cite{Jia2014, PhysRevLett.110.265502}. Figs.~\ref{fig2}(c) and (d) show the calculated spectra, which capture the two- and four-spinon excitations at both edges.

\begin{figure}[tb]
	\begin{center}
		\includegraphics[width=89mm]{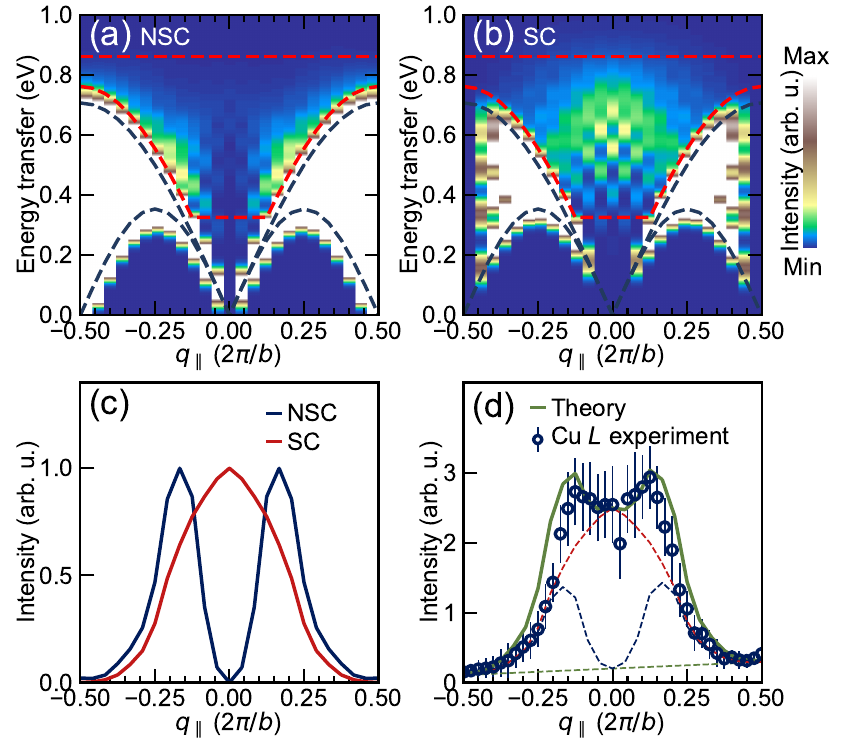}
		\caption{The calculated Cu $L_3$-edge RIXS spectra in the (a) NSC and (b) SC channels, computed using the Kramers-Heisenberg formalism. The white colour represents the maximum spectral weight in the colour scale. (c) the normalised intensity profiles of the two spin channels obtained by integrating the region marked by the red dashed lines in (b) and (c). (d) the integrated intensity profile for the same region obtained from the Cu $L_3$-edge RIXS experiment on SrCuO$_2$. The continuous line is a weighted sum of the NSC and SC channel profiles shown in panel (c) and a linear background to match the experimental profile (see text).}
		\label{fig3}
	\end{center}
	\vspace{-0.75cm}
\end{figure}

The RIXS spectra at the Cu $L_3$-edge has contributions from both the NSC and SC channels~\cite{PhysRevLett.103.117003, PhysRevX.6.021020} while the O $K$-edge has contributions from only the SC channel. To demonstrate this, Fig.~\ref{fig3} shows the spin-resolved contributions to the Cu $L_3$-edge RIXS intensity from the (a) NSC and (b) SC channels, highlighting the features outside the two-spinon phase space.  Panel (c) shows the computed intensity profiles of the two channels for momentum transfers along the chain, obtained by integrating the spectral weight along the energy axis within the region enclosed by the red dashed lines in panels (a) and (b). The profiles have been normalised to their maximum values in the probed portion of the Brillouin zone for comparison. From panels (a) and (c), it is clear that the four-spinon excitations in the NSC channel have the maximum spectral weight close to the upper boundary of the two-spinon continuum and a negligible weight at zone center. Conversely, the four-spinon excitations observed in the SC channel have maximal spectral weight close to $q_{\parallel}=0$. In panel (d), we show the experimental Cu $L_3$-edge RIXS intensity profile from the  same region (see Fig.~\ref{fig2}(b) and Supplementary Materials Fig.~S2 for the full map)~\cite{SM}. The experimental profile can be well described by a sum of the NSC and SC channel profiles (in a ratio $\sim1:1.92$) and a linear background. Unlike INS, our results show that both NSC and SC four-spinons have significant presence outside the two-spinon phase space when probed using Cu $L_3$-edge RIXS. This result also suggests why the spectral weight distribution in this region of phase space is substantially different from that at O $K$-edge, where only SC excitations are allowed. 

\begin{figure}[tbp]
	\begin{center}
		\includegraphics[width=89mm]{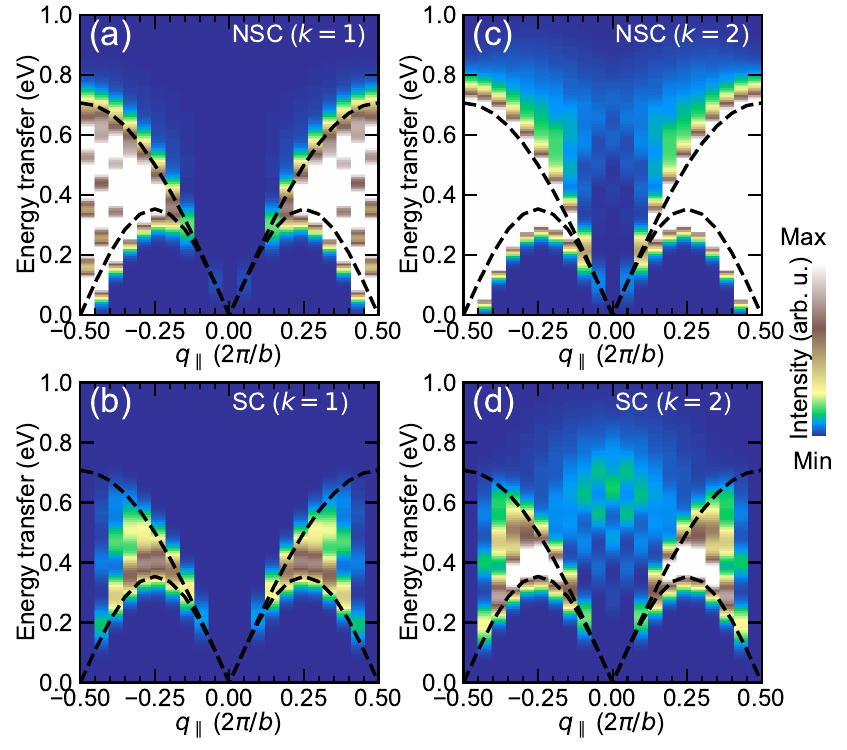}
		\caption{The generalized dynamical structure factors $S^k_\mathrm{(N)SC} (q, \omega)$ appearing in higher-order terms of the UCL expansion. (a) and (b) show the intensity maps of the first-order ($k=1$) terms in the NSC and SC channels, respectively.  (c) and (d) show the intensity maps of the second-order ($k=2$) terms in the NSC and SC channels, respectively. The white colour represents the maximum spectral weight in the colour scale.}
		\label{CorrelationFunctions}
	\end{center}
	\vspace{-0.75cm}
\end{figure}

While the KH formalism captures the spectral weight distribution of both the two- and four-spinon excitations, it is desirable to identify the correlation functions leading to these excitations using the UCL expansion~\cite{PhysRevB.75.115118,PhysRevB.77.134428,PhysRevX.6.021020}. For a half-filled antiferromagnetic $d^9$ cuprate, the zeroth-order term in the expansion of the SC and NSC channels are elastic scattering and the conventional dynamical spin structure factor $S(q,\omega)$,  respectively~\cite{SM,PhysRevX.6.021020,UKUMAR2019}. Corrections to these terms are provided by the higher-order terms in the UCL expansion. The next-order generalized dynamical structure factors in the NSC and SC channels are shown in Fig.~\ref{CorrelationFunctions}. The spectra at $k^{th}$ order are 
\begin{equation*}
	S^k_\mathrm{(N)SC} (q, \omega) = \sum_{f}\left\vert \langle f|O^\mathrm{(N)SC}_{q,k}|g\rangle \right\vert^2\delta\left(E_{f} - E_{g} - \omega\right),
\end{equation*}
where $k$ indexes the $k^\text{th}$ order in the UCL approximation, 
$O^\mathrm{(N)SC}_{q,k} = \frac{1}{\sqrt{N}}\sum_{i} e^{i q R_i} O_{i,k}^\mathrm{(N)SC}$, 
and $O_{i,k}^\mathrm{NSC} = S_i^z \left(\textbf{S}_i\cdot (\textbf{S}_{i+1}+\textbf{S}_{i-1})\right)^k $ and $O_{i,k}^\mathrm{SC} = \left(\textbf{S}_i\cdot (\textbf{S}_{i+1}+\textbf{S}_{i-1}) \right)^k $ are the effective operators in each channel. The results in  Figs.~\ref{CorrelationFunctions}(a) and (b) demonstrate that the spectral weight remains confined within the two-spinon phase space~\cite{PhysRevB.83.245133, PhysRevLett.106.157205} for the first-order corrections to both the NSC and SC channels. 
Spectral weight outside the two-spinon phase space in either channel appears only with the second-order corrections [Fig.~\ref{CorrelationFunctions}(c) and (d)]. The four-spinons seen in Cu $L_3$-edge RIXS experiment are, therefore, beyond the scope of first order nearest-neighbour correlation functions and represent higher order and longer-range spin correlations (see SM, \cite{SM}).

In conclusion, we have demonstrated that both O $K$- and Cu $L_3$-edge RIXS can probe four-spinon excitations outside the two-spinon phase space in the low-dimensional quantum magnet SrCuO$_2$. We further showed that these four-spinon excitations involve long-range spin-flips due to the lifetime of the intermediate state's core-hole. Higher-order corrections in the UCL expansion in the SC channel have been explored previously in the context of two-magnon excitations in Cu $K$-edge RIXS studies of two-dimensional cuprates~\cite{PhysRevB.77.134428, PhysRevLett.100.097001}. But the relevance of these corrections for the Cu $L_3$-edge RIXS has not been widely explored~\cite{PhysRevX.6.021020}. Unexpectedly, we find that corrections up to the second-order in the UCL expansion are needed for both the SC and NSC channels to achieve even a qualitative description of the four-spinon excitations at Cu $L_3$-edge. Our work illustrates that the Cu $L_3$-edge RIXS is more sensitive to long-range quantum spin fluctuations in low-dimensional quantum magnets than previously thought. In particular, the NSC channel at this edge probes completely new long-range correlation functions, which one can exploit for exploring higher-order many-body spin dynamics.

\section*{Acknowledgements}
U.~K. acknowledges support from U.S. DOE NNSA under Contract No. 89233218CNA000001 through the LDRD Program. S.~J. acknowledges support from the National Science Foundation under Grant No. DMR-1842056. T. S. was supported by the Swiss National Science Foundation (SNSF) through the SNSF research grants 200021$\_$178867 and 200021L$\_$141325. We acknowledge Diamond Light Source for providing the beam time on I21 beam line under Proposal No. NR21184. We acknowledge T. Rice for the technical support throughout the beam time. We also thank G. B. G. Stenning and D. W. Nye for help on the Laue instrument in the Materials Characterisation Laboratory at the ISIS Neutron and Muon Source. 

\bibliography{SrCuO2ref}

\makeatletter

\pagebreak
\widetext
\begin{center}
	\textbf{\large Supplemental Materials: Unraveling higher-order corrections in the spin dynamics of RIXS spectra}
\end{center}
\setcounter{equation}{0}
\setcounter{figure}{0}
\setcounter{table}{0}
\setcounter{page}{1}
\makeatletter
\renewcommand{\theequation}{S\arabic{equation}}
\renewcommand{\thefigure}{S\arabic{figure}}
\bibliographystyle{apsrev4-1}
\setcounter{section}{0}
\renewcommand{\thesection}{S-\Roman{section}}

\section{Experimental details}
\begin{figure}[h]
	\centering
	\includegraphics[width=0.5\linewidth]{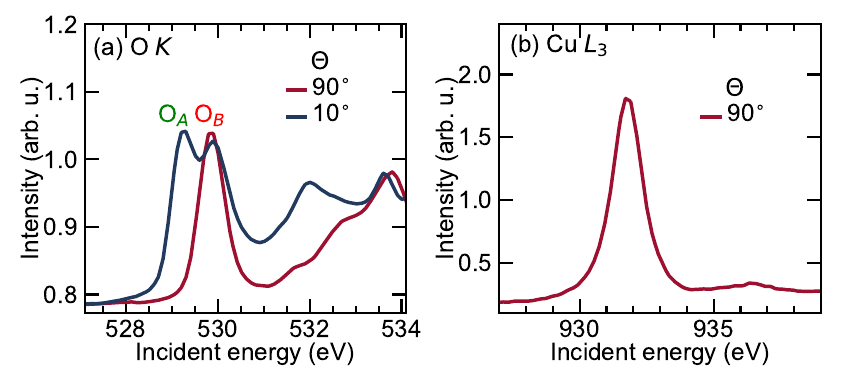}
	\caption{Determination of the resonant X-ray energies corresponding to O$_B$ and Cu sites for SrCuO$_2$ shown in Fig.~1(a) of main text. (a) O $K$-edge X-ray absorption (XAS) with $\pi$ polarisation along ($90^{\circ}$) and normal to the spin chain direction  ($10^{\circ}$), resonating with O$_B$ and O$_A$ sites, respectively. (b) Cu $L_3$-edge XAS with $\pi$ polarisation along ($90^{\circ}$) the spin chain direction, resonating with Cu site.}
	\label{fig:SIFig_XAS.pdf}
\end{figure}

\begin{figure}[h]
	\centering
	\includegraphics[width=0.5\linewidth]{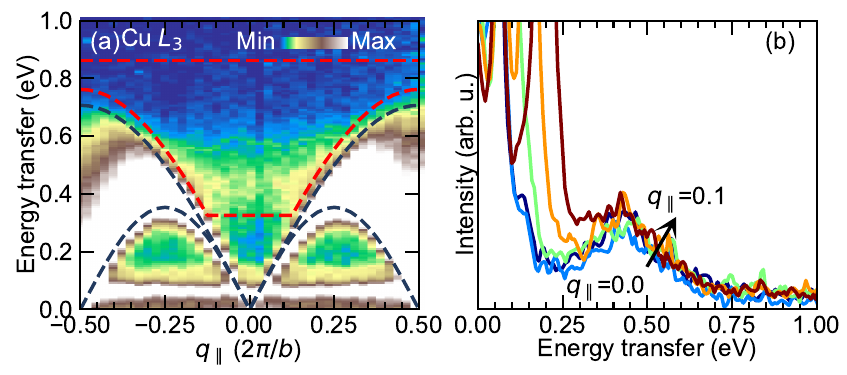}
	\caption{(a) Full Cu-$L_3$-edge RIXS intensity map for SrCuO$_2$. The two-spinon continuum boundaries are shown by blue dashed lines.  In Fig.~3(d) of main text, the  integrated intensity profile for the region  marked  by  the  red  dashed lines is shown. (b) Cu $L_3$-edge RIXS line spectra close to $q_{\parallel}$=0 highlighting the flat four-spinon spectral weights.}
	\label{fig:SIFig_fullmap}
\end{figure}

\section{Ultra-short core-hole lifetime expansion of Kramers-Heisenberg Formalism at Cu $L_3$-edge}
The Kramers-Heisenberg (KH) expression for the RIXS intensity can be re-expressed as a series of ``simpler'' correlation functions using the ultra-short core-hole lifetime (UCL) approximation~\cite{PhysRevB.77.134428, PhysRevX.6.021020}. The UCL approximation at the Cu $L_3$-edge should be a good approximation to the RIXS intensity, as long as $J/\Gamma < 1$. 

The RIXS intensity in the KH formalism is given by 
\begin{equation}
I(q,\omega) \propto \sum\limits_{f}\left\lvert
\langle f \lvert D^{\dagger}_{k_\mathrm{out}} \mathcal{O}  D^{\phantom\dagger}_{k_\mathrm{in}} \rvert g \rangle
\right\rvert^{2}\delta\left(E_{f} - E_{g} - \omega\right),
\end{equation}
where 
\begin{equation}
\mathcal{O} = \frac{1}{\omega_\text{in}-\mathcal{H}+i\Gamma}  = \sum_{|N\rangle}\frac{\vert N\rangle\langle N\vert }{\omega_\text{in}-\mathcal{E}_N+i\Gamma}.
\end{equation}
Here, $\vert N\rangle$ are the eigenstates of intermediate Hamiltonian, $\mathcal{H} = \tilde{H} + H_\lambda$, where $\tilde{H} = H_0+U_c\sum_{i} n_i n_i^c $, $H_0 = J\sum_i \textbf{S}_i\cdot \textbf{S}_{i+1}$ and $H_\lambda = \epsilon_c^\alpha \sum_{i\alpha} n_{i\alpha}^c$, $n_i$ and $n_i^c$ are occupation in the valence and core-level respectively, and $U_c$ is the Coulomb interaction between the two levels. In $n_{i\alpha}^c$, $\alpha \in \{L_2, L_3\}$ and characterizes the spin-orbit coupling in the core-level with energy $\epsilon_c^\alpha$. We work in the hole language throughout this derivation.  Also, we have ignored the hopping term in $H_0$ since we are interested only in the half-filled case and it should not contribute significantly to the intermediate state in the limit of large core-hole potential as the charges are immobile. 

Due to the large $U_c$ on the core-hole site, hopping to the $d$-orbital is largely prohibited in the intermediate state. Assuming that hopping to this site is completely suppressed, we can decompose the intermediate states as $ \vert N\rangle =\vert n\rangle \vert n_c\rangle,~ \mathcal{E}_n = E_n +E_c $. Here, $\vert N\rangle~(\mathcal{E}_N)$, $\vert n\rangle~(E_n)$ and $\vert n_c\rangle~(E_c)$ are the eigenstates~(energies) of the $\mathcal{H}$, $\tilde{H}$ and $H_\lambda$ respectively. One can rewrite the operator $\mathcal{O}$ in its spectral representation as 
\begin{equation}
\frac{1}{\omega_\text{in}-\mathcal{H}+i\Gamma} = \sum_{|n,n_c\rangle}\frac{\vert n\rangle\langle n\vert n_c\rangle\langle n_c\vert }{\omega_\text{in}-E_n -E_c+i\Gamma}.
\end{equation}

$E_c$ can be tuned to either the $L_{2}$ ($E_{L_2}$) or $L_3$ resonances ($E_{L_3}$), which are well separated in energy due to large spin-orbit coupling in the core $2p$-level of the Cu atom. In our discussion, we discuss the simplification in the context of $L_3$ edge. Though it can be easily extended to $L_2$-edge by replacing  $L_2$ with $L_3$ and considering the relevant approximations. For $L_3$-edge, we have
\begin{equation}
\begin{aligned}
\mathcal{O} = \sum_{|n\rangle}\frac{\vert n\rangle\langle n \vert L_3\rangle\langle L_3\vert }{\omega_\text{in}-E_n -E_{L_3}+i\Gamma}. 
\end{aligned}
\end{equation}
It is important to note that $|L_3\rangle$ can either flip or conserve the spin in core-hole and hence lead to non-spin-conserving (NSC) $\Delta S_{\mathrm{tot}}=1$ and spin-conserving (SC) $\Delta S_{\mathrm{tot}}=0$ channels, respectively. We define $\omega_\text{in}-E_{L_3} = \Delta$ 
such that 
\begin{equation}
\begin{aligned}
\sum_{|n\rangle}\frac{\vert n\rangle\langle n\vert L_3\rangle\langle L_3\vert }{\omega_\text{in}-E_n -E_{L_3}+i\Gamma} & = |L_3\rangle\langle L_3| \frac{1}{\Delta-\tilde{H}+i \Gamma} = \\
& =|L_3\rangle\langle L_3|\frac{1}{\Delta+i\Gamma}\sum_{l=0}^{\infty}\bigg( \frac{\tilde{H}}{\Delta+i\Gamma}\bigg)^l.
\end{aligned}
\end{equation}






\noindent\textit{Resonance condition and UCL approximation} --- 
When the incident energy is tuned to resonate with the $L_3$-edge such that $\Delta\rightarrow 0$, the propagator in the above expression can be written as
\begin{equation}
\begin{split}
\frac{1}{i\Gamma-\tilde{H}} 
=\frac{1}{i\Gamma}\sum_{l=0}^{\infty}\bigg( \frac{\tilde{H}}{i\Gamma}\bigg)^l.
\end{split}
\end{equation}

We now consider the application of the dipole operator at the Cu $L_3$ edge $D_{k_\text{in}} = \sum_{i, \sigma} e^{i k R_i} D_{i,\sigma}$, where $D_{i,\sigma} = \sum_{\alpha} d^{\phantom\dagger}_{i \sigma}  p_{i\alpha}^\dagger$ annihilates a hole in the $d$-orbital at site $i$. In this case, $H_{i} D_{i,\sigma}|g\rangle = 0 $, where $H_i =J~\textbf{S}_i\cdot (\textbf{S}_{i+1} + \textbf{S}_{i-1})$. Since the core-hole potential term $H_c$ inhibits hopping to site $i$ and superexchange coupling to the spin at the core-hole site, we can write the intermediate state Hamiltonian as $\tilde{H} = H_0 - H_i$. One can further see that $H_0 -H_{i}$ commutes with $D_{i,\sigma}$ and we have $\tilde{H}|g\rangle =(H_0+H_c)|g\rangle= 0|g\rangle$. (Here, we have set ground state energy, $H_0 |g\rangle = 0|g\rangle$.) 
Using these properties, one can write 
\begin{equation}
\begin{aligned}
\sum_{l=0}^{\infty}\frac{\tilde{H}^l}{(i\Gamma)^{l+1}} D_{i,\sigma} |g\rangle &= \sum_{l=0}^{\infty}D_{i,\sigma} \frac{(H_0-H_i)^l}{(i\Gamma)^{l+1}}|g\rangle\\
&= D_{i,\sigma} \frac{1}{i\Gamma}\bigg(1 - \frac{H_i}{i\Gamma} + \frac{H_i^2}{(i\Gamma)^2}-\frac{H_0 H_i}{(i\Gamma)^2} +\dots \bigg) |g\rangle 
\end{aligned}
\end{equation}
Now, one can see that RIXS cross-section can be written in terms of correlation functions involving powers of $H_i$. 

Using the relation, $ d_{i \sigma}^\dagger d^{\phantom\dagger}_{i \sigma} p^{\phantom\dagger}_{i\alpha}  |L_3\rangle \langle L_3|  p_{i\alpha}^\dagger = n_{i\sigma}$ and $ d_{i \bar{\sigma}}^\dagger d^{\phantom\dagger}_{i \sigma} p^{\phantom\dagger}_{i\alpha}  |L_3\rangle \langle L_3|  p_{i\alpha}^\dagger = S_i^z$~\cite{PhysRevLett.103.117003, PhysRevB.85.064423}, the scattering amplitude with the core-hole term in the expansion can be written as
\begin{equation}
\begin{aligned}
\langle f| e^{i\textbf{q}\cdot\textbf{R}_i} D_{i,\sigma'}^\dagger |L_3\rangle\langle & L_3|D_{i,\sigma} \sum_{l=0}^{\infty}e^{i\textbf{q}\cdot\textbf{R}_i}\frac{(H_0-H_i)^l}{(i\Gamma)^{l+1}} |g\rangle       =
\begin{cases}
\langle f|\sum_{ i }e^{i\textbf{q}\cdot\textbf{R}_i} n_{i,\sigma} \sum_{l=0}^{\infty}\frac{(H_0-H_i)^l}{(i\Gamma)^{l+1}} |g\rangle & \text{if $\sigma' = \sigma$}   \\
\langle f|\sum_{ i }e^{i\textbf{q}\cdot\textbf{R}_i} S_{i}^z \sum_{l=0}^{\infty}\frac{(H_0-H_i)^l}{(i\Gamma)^{l+1}} |g\rangle & \text{if $\sigma' \neq \sigma$}.
\end{cases}
\end{aligned}
\end{equation}
Neglecting the cross terms, the SC and NSC channels with leading correction terms corresponding to $H_i$ and $H_i^2$ are given by 
\begin{equation}
\begin{aligned}
\mathcal{I}_{SC}^{UCL}(\textbf{q},\omega) =& \Bigg( \frac{1}{\Gamma^2} \sum_f \Big|\langle f|\sum_{ i }e^{i\textbf{q}\cdot\textbf{R}_i} n_{i,\sigma} |g\rangle \Big|^2+\frac{J^2}{\Gamma^4}\sum_f \Big|\langle f|\sum_{ i }e^{i\textbf{q}\cdot\textbf{R}_i} \textbf{S}_{i}\cdot(\textbf{S}_{i+1}+\textbf{S}_{i-1}) |g\rangle \Big|^2 \\ & +\frac{J^4}{\Gamma^6} \sum_f \Big|\langle f|\sum_{ i }e^{i\textbf{q}\cdot\textbf{R}_i} \big( \textbf{S}_{i}\cdot(\textbf{S}_{i+1}+\textbf{S}_{i-1})\big)^2 |g\rangle \Big|^2 + \dots
\Bigg) \delta(\Omega+E_g-E_f)
\end{aligned}
\end{equation}
In the case of SC-channel, the leading term corresponds to elastic scattering at half-filling. The first order correction is the spin-exchange dynamical factor $S^\text{exch}(\textbf{q}, \omega)$ \cite{PhysRevLett.106.157205}, which is important for  indirect RIXS. Similarly for the spin flip channel,  
\begin{equation}
\begin{aligned}
\mathcal{I}_{NSC}^{UCL}(\textbf{q},\omega) =& \Bigg( \frac{1}{\Gamma^2} \sum_f \Big|\langle f|\sum_{ i }e^{i\textbf{q}\cdot\textbf{R}_i} S_{i}^{z} |g\rangle \Big|^2+\frac{J^2}{\Gamma^4} \sum_f \Big|\langle f|\sum_{ i }e^{i\textbf{q}\cdot\textbf{R}_i} S_{i}^{z}\textbf{S}_{i}\cdot(\textbf{S}_{i+1}+\textbf{S}_{i-1}) |g\rangle \Big|^2 \\ & +\frac{J^4}{\Gamma^6} \sum_f \Big|\langle f|\sum_{ i }e^{i\textbf{q}\cdot\textbf{R}_i} S_{i}^{z}\big(\textbf{S}_{i}\cdot(\textbf{S}_{i+1}+\textbf{S}_{i-1})\big)^2 |g\rangle \Big|^2 + \dots \Bigg) \delta(\Omega+E_g-E_f).
\end{aligned}
\end{equation}

The response function for first and second order correction is presented in the Fig.~4 of the main text.\\ 



\noindent{\sl Simplification of Second Order} ---
The second order correction can further be simplified 
\begin{equation}
\begin{aligned}
H_i^2/J^2 = & \big(\textbf{S}_i\cdot(\textbf{S}_{i-1}+\textbf{S}_{i+1})\big)\big(\textbf{S}_i\cdot(\textbf{S}_{i-1}+\textbf{S}_{i+1})\big)\\
=& (\textbf{S}_i\cdot\textbf{S}_{i-1})^2+(\textbf{S}_i\cdot\textbf{S}_{i+1})^2+ (\textbf{S}_i\cdot\textbf{S}_{i-1})(\textbf{S}_i\cdot\textbf{S}_{i+1}) 
+(\textbf{S}_i\cdot\textbf{S}_{i+1})(\textbf{S}_i\cdot\textbf{S}_{i-1}).
\end{aligned}
\end{equation}


Notice that only the last two terms can contribute to the longer-range correlation functions. Using the vector triple product and identity reported, one can rephrase the above operators as 
$(\textbf{S}_i\cdot\textbf{S}_{i+1}) (\textbf{S}_i\cdot\textbf{S}_{i-1}) 
= (\textbf{S}_{i-1}\cdot \textbf{S}_{i+1})\textbf{S}_i^2 - \textbf{S}_{i+1}\times (\textbf{S}_{i}\times\textbf{S}_{i-1})\cdot \textbf{S}_i$
Notice that in the first term $\langle\textbf{S}_i^2\rangle  = \frac{3}{4}$ as we have a spin-1/2 system. It is also consistent with the Eq.(37) of Ref.~\cite{PhysRevB.77.134428}, where long-range correlation functions for the SC channel were reported. Longer-range correlation functions has also be proposed and explored in Ref.~\cite{PhysRevB.85.064422} in the context of 1D systems.  From the first term in the vector triple product, it is clear that the second-order correction has a term with the long-range correlation with the form
\begin{equation}\label{UCLsecondorder}
\begin{aligned} 
H_i^2 \propto J^2\textbf{S}_{i-1}\cdot\textbf{S}_{i+1}.
\end{aligned}
\end{equation}
\begin{figure}[t]
\centering
\includegraphics[width=0.5\linewidth]{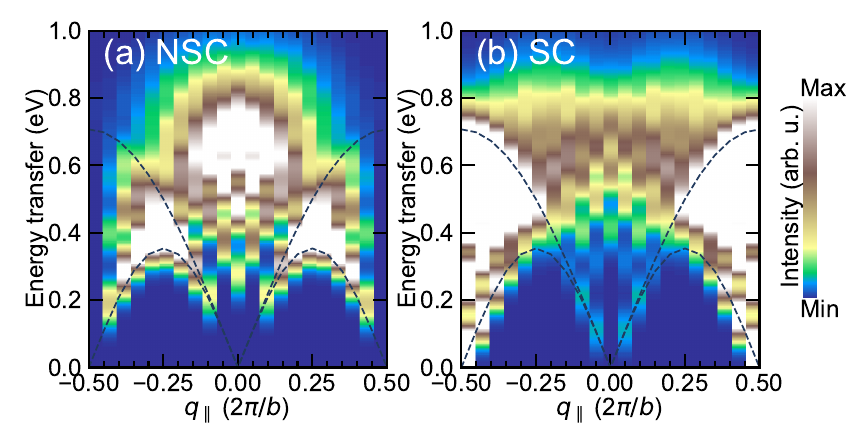}
\caption{ $S_\mathrm{(N)SC}^{l=2} (q, \omega)$ response. (a) and (b) show the result for non-spin conserving and spin conserving channel result, respectively, for the next-nearest neighbour spin-flip operators given by Eq.~\eqref{Simp_SecondOrder} that arise in the second order term of the UCL approximation.} 
\label{fig:P2correlation}
\end{figure}

We plot the next nearest-neighbor response given by this term in Fig.~\ref{fig:P2correlation}, where the correlation operator is given by
\begin{equation} \label{Simp_SecondOrder}
S_\mathrm{(N)SC} (q, \omega) = \sum_{f}\left\vert \langle f|O^\mathrm{(N)SC}_{q}|g\rangle \right\vert^2\delta\left(E_{f} - E_{g} - \omega\right),
\end{equation}
where $O^\mathrm{(N)SC}_{q} = \frac{1}{\sqrt{N}}\sum_{i} e^{i q R_i} O_{i}^\mathrm{(N)SC}$, 
and $O_{i}^\mathrm{NSC} = S_i^z \left(\textbf{S}_{i-1}\cdot \textbf{S}_{i+1}\right) $ and $O_{i}^\mathrm{SC} = \left(\textbf{S}_{i-1}\cdot \textbf{S}_{i+1}\right)$ are the effective operators in each channel. The above correlation function generates excitations in the new phase space, as shown in the Fig.~\ref{fig:P2correlation}.

\makeatother
\end{document}